# *A Study on Silicon Nanotubes based on the Tersoff potential*

**Jeong Won Kang*†, Jae Jeong Seo‡, and Ho Jung Hwang†‡**

† Institute of Technology and Science, Engineering Building, Chung-Ang University, 221 HukSuk-Dong, DongJak-Ku, Seoul 156-756, Korea

‡ Computational Semiconductor Laboratory, School of Electric and Electronic Engineering, Chung-Ang University, 221 HukSuk-Dong, DongJak-Ku, Seoul 156-756, Korea

This study showed the structures and the thermal behaviors of Si nanocages and nanotubes using classical molecular dynamics simulations based on the Tersoff potential. For hypothetical Si nanotubes based on the Tersoff potential, Si-Si bond length, cohesive energies per atom, diameters, and elastic energy to curve the sheet into tube were in good agreement with those obtained from previous density-functional theory results. Most of the structures, which were obtained from the SA simulations for several initial structures with diamond structure, have included encaged, tubular, or sheet-like structures and have been composed of both $sp^3$ and $sp^2$ bonds. The cohesive energies per atom for silicon nanotubes were higher than that for the Si bulk in the diamond structure, and this implies the difficulty in producing silicon nanotubes or graphitelike sheets. However, since the elastic energy per atom to curve the sheet into tube for silicon atoms is very low, when graphitelike sheets of silicon are formed, the extra cost to produce silicon nanotubes is also very low. When silicon nanotubes are composed of both $sp^2$ and $sp^3$ bonds and the ratio of $sp^3$ to $sp^2$ is high, since the probability of the existence of silicon nanotubes increases, silicon nanotubes similar to multi walled structures are anticipated.

# 1. Introduction

Discovery of carbon nanotubes (CNT) as a by-product of fullerene synthesis by Iijima [1,2] in the early 1990s opened a challenging new field in nano-scale materials. Although CNTs can be readily obtained using a variety of techniques, such as arc-discharge [3], laser-ablation [4], etc., synthesis of nanotubes having other chemical compositions has only been reported in a relatively small number of cases. Examples are $B_xC_yN_z$ composite nanotubes [5-9], Mo and W chalcogenide tubes [10-12], and NiCl cage structures and nanotubes [13]. Futrhermore, theoretical predictions of the stability and the electronic structure of GaN [14], GaSe [15], and P [16] nanotubes have been reported in the literatures.

New forms of carbon such as fullerenes and nanotubes have been providing increasing challenges for silicon in the field of nanotechnology. However, since silicon-based materials have been the focus of extensive research due to their technological importance, new forms of stable silicon are required to sustain the current silicon-based technology. Recently, ultrathin silicon nanowires have aroused growing interest in nanoscience and nanotechnology as possible elements of nanoelectronic devices; for example, the prototypes of CNT and silicon nanowires heterojunctions could be realized [17] and many theoretical studies on ultrathin silicon nanowires have been done using atomic simulations. Menon and Richter [18] proposed quasi-one-dimensional silicon structure, characterized by a core of bulklike fourfold-coordinated atoms surrounded by a structure related to well-known reconstructed surfaces with a large part of threefold-coordinated atoms. Marsen and Sattler [19] proposed a model consisting of fullerenelike structures for silicon nanowires. Seifert *et al* [20] proposed silicon based tubular nanostructures, silicide nanotubes. Li *et al* [21] proposed thin short nanowires consisting of tricapped and uncapped trigonal prisms, and showed the structure of a silicon nanotube (SiNT) with five tricapped trigonal prisms corresponding to armchair (3, 3) CNT. Fagan *et al* [22,23] investigated the electronic, the structural and the thermal properties of three hypothetical SiNTs with the structures of CNTs. Ivanovskaya *et al* [24] investigated hypothetical SiNTs containing regular chains of metallocarbohedrenes using one dimensional tight-binding model within Hűckel approximation. It should be noted that even though both silicon and carbon are isovalent, their behavior in forming chemical bonds is quite different. For example, the $sp^2$ hybridization is more stable in carbon, whereas the $sp^3$ hybridization is more stable in silicon. Therefore, carbon easily forms graphite, fullerene, and nanotube composed of only $sp^2$ bonds, whereas silicon has been well-known in the form of only diamond structure. However, we can't completely rule out the possibility of the existence of SiNTs. Bahel and Ramakrishna [25] revealed that the lowest energy structure of the $Si_{12}$ cluster is a bicapped pentagonal antiprism, called a hollow icosahedron. Recently, new endohedral silicon cage clusters were successfully synthesized in the gas phase [26], using transition metal atoms with a partially filled *d* shell as aggregation centers for silicon atoms. A SiNT was stabilized by an encapsulation of the Ni chain using tight-binding molecular dynamics and *ab initio* method [27]. Fagan *et al* [23] showed that there is a significant cost to produce graphitelike sheets of silicon, but once they are formed, the extra cost to produce the tubes is of the lower cost than that in carbon.

In this paper, we present the results of classical molecular dynamics (MD) simulations based on the Tersoff potential [28] and discuss about the possible stability of some hypothetical SiNTs and a $Si_{60}$ fullerene.

## 2. Empirical potential

For the Si-Si interactions, we used a many-body empirical potential, the Tersoff potential [28]. Tersoff [28,29] has shown that, if this function is used for the pair terms, a wide range of structural properties of materials, including carbon and silicon, can be appropriately described using the Tersoff potential, providing a reasonable starting point for predicting trends, such as those studied here. This empirical potential was fit to the lattice constant and binding energy of a number of silicon lattices as well as the elastic constants and vacancy formation energies of diamond structure. This potential was used by Robertson *et al* [30] to study the energetic and elastic properties of CNT and by Hamada *et al* [31] to generate tube structures, subsequently used in tight-binding electronic structure calculation.

Optimal atomic configurations of hypothetical SiNTs and $Si_{60}$ were obtained using the steepest descent (SD) method, which is the simplest of the gradient methods, from the atomic configurations of CNTs having C-C bond length, 1.42 Å. The choice of direction was determined by where the force exerted by interatomic interaction decreased the fastest, which was in the opposite direction to $\nabla E_i$, where $E_i$ is the potential energy of *i*th atom. In this work, the SD method was applied to the atomic positions, and the next atomic position vector ($\boldsymbol{r}'_i$) was obtained by a small displacement of the existing atomic position vector ($\boldsymbol{r}_i$) along a chosen direction under the condition, $|\boldsymbol{r}'_i - \boldsymbol{r}_i|/|\nabla E_i| = 0.001$.

In the case of the optimal structures of the hypothetical zigzag (10, 0) SiNTs by the SD method, the optimal diameter, 12.861 Å, based on the Tersoff potential is in good agreement with the density-functional theory (DFT) result, 12.41 Å [22,23]. For the Tersoff potential, the optimal Si-Si bond length and the potential energy per atom are 2.305 Å and –3.899 eV, respectively. According to the DFT result [22], The Si-Si bond length and the potential energy per atom were 2.245 Å and –3.83 eV, respectively. The value for the potential energy per atom obtained from the Tersoff potential is 0.731 eV/atom higher than that for the diamondlike structure. Considering that the cohesive energy for the Si bulk in the diamond structure is 4.63 eV/atom, the cohesive energies for the studied nanotubes are only 84.21 % of the bulk, similar to 82 % obtained from the DFT result [23]. Comparing with CNTs that have been around 99 % of the cohesive energy that they would have in perfect crystalline, we have a clear understanding of the difficulty in producing SiNTs. For carbon, the energy cost for curving the sheet into a cylinder for (10, 10) nanotube has been known to be only 0.05 eV/atom [39]. A total energy value for the graphite sheet of silicon sheet is 0.719 eV/atom higher than that for the silicon in the diamond structure. Therefore, we obtain the order of 0.731 - 0.719 = 0.012 eV/atom to curve the sheet into (10, 0) SiNT. This value for (10, 0) SiNT is 0.05 eV/atom in the Tersoff potential and was 0.04 eV/atom in the DFT result [23], respectively. Fagan *et al* [23] also presented a systematic study on the thermal behavior of the hypothetical SiNTs using the Tersoff potential. Therefore, considering above results, we think that the results of hypothetical SiNTs using the Tersoff potential are in good agreement with those using the DFT and the Tersoff potential can be efficiently applied to the investigation of Si nanocages and nanotubes.

### 3. Simulation Procedures

Our MD simulations used the same MD method as in our previous works [32-37], with time step of 0.5 fs. The MD code used the velocity Verlet algorithm, a Gunsteren–Berendsen thermostat to maintain constant temperature, a periodic boundary condition (PBC), and neighbor lists to improve the computing performance [38].

To obtain the structures of Si nanowire, we have used a simulated annealing (SA) method that has been applied to the initial structures in Table 1. After MD simulations for the initial structures in Table 1 have been performed during 10 ps at high temperature, the kinetic temperatures of systems have been decreased by the quenching rate 1 or 5 % by 5 ps to 10 K. Initial structures consist of three orientations, {111}, {110} and {100}, and each orientation also has three diameters. Table 1 shows the initial structures, diameters, number of atom, and length of the PBC.

On heating of hypothetical SiNTs, the kinetic temperature increased from 300 K by 50 K interval. At each temperature, MD runs of $2\cdot10^5$ steps were made with a time step of 0.5 fs (total 100 ps) and the statistical data were obtained from the last $10^3$ steps. On heating of $Si_{60}$, the kinetic temperature increased by the heating rate 5 % from 10 to

400 K and by 20 K interval from 400 to 1200 K.

## 4. Results and discussion

Figure 1 shows the variation of the cohesive energy per atom as a function of temperature for the SA simulations of four systems with the quenching rate 5 % by 5 ps. In Fig. 1, the curves obtained from the SA simulations are linear regions below 500 K, whereas the curve for $Si_{60}$ fullerene on heating is almost linear region below 950 K. The linear regions of the curves in Fig. 1 mean that a definite structure maintains. Cage-like $Si_{52}$ as shown in Fig. 2(a) was obtained from the SA simulation with the quenching rate 5 % by 5 ps from 850 K for A1 case without the PBC, and the entrance of $Si_{52}$ cage is composed of 9 Si atoms as shown in Fig. 2(a). Figure 2(b) shows a cage structure composed of 32 Si atoms that were extracted from the final structure obtained from the SA simulation with the quenching rate 5 % by 5 ps from 1000 K for A1 case with the PBC. The entrance of $Si_{32}$ cage is composed of 8 Si atoms as shown in Fig. 2(a).

In this work, $sp^3$, $sp^2$ and $sp^1$ represent fourfold, threefold and twofold coordinated atoms, respectively. Most of the final structures, which were obtained from the SA simulations for the initial structures in Table 1, have included encaged, tubular, or sheet-like structures composed of pentagon, hexagon, heptagon, etc. They have been composed of both $sp^3$ and $sp^2$ bonds, as shown in Figs. 2-4, 6 and 7. However, most of them do not have well-defined structures but have some distorted regions or several defects as shown in Figs. 2-4. A few cases have shown well-defined structures as shown in Figs. 6 and 7. Figure 3(a) shows the final structure of A2 case that includes a tubular and a few encaged structures, and the circle indicates the tubular structure. Figure 3(b) shows the sheet-like structure of the outside wall of the tubular structure indicated by the circle in Fig. 3(a). Figure 4 shows the final structure of A2 case which was obtained from the SA simulation with quenching rate 1 % by 5 ps from 1200 K, and most of this structure are similar to a spreading sheet composed of pentagon, hexagon, heptagon, etc. Since the SA simulation of Fig. 4 case started at higher temperature than that of Fig. 3 case, Si atoms in Fig. 4 case were spread more widely than Si atoms in Fig. 3 case, and then the structure of Fig. 4 became a spreading sheet-like structure.

The cohesive energies per atom for hypothetical SiNTs are higher than that for the Si bulk in the diamond structure, and this implies the difficulty in producing SiNTs or Si graphitelike sheets. However, since the elastic energy per atom to curve the sheet into cylinder for Si atoms is low as much as that for carbon atoms, if graphitelike sheets of Si are formed, the extra cost to produce the tubes is of the similar order of that in carbon. Therefore, in our simulations, the final structures including graphitelike structures also shows encaged or tubular structures.

Figure 5 shows structures of $Si_{60}$ fullerene for temperature. $Si_{60}$ fullerene maintained its original ball-like structure below 1000 K. At the collapse temperature, the cohesive energy curve of $Si_{60}$ fullerene in Fig. 1 shows the obvious upward curvature. This means that $Si_{60}$ fullerene based on the Tersoff potential is a stable structure in the condition of low temperature.

Above results shows the fullerenelike Si cages. Mitas *et al* [39] showed a $Si_{20}$ cage cluster by the DFT-quantum Monte Carlo (QMC) electronic structure and Li *et al* [21] showed a capped $Si_{42}$ cage structure by the full-potential (FP) linear-muffintin-orbit (LMTO) MD simulation. Marsen and Sattler [19] assembled ultrathin Si nanowire bundles from a magnetron sputter source and the nanowires were from 3 to 7 nm in diameter. In order to understand the observed quasi-one-dimensional structures, they constructed the diamondlike and fullerenelike wire models, obtained the binding energies and the band gaps of such structures by molecular-orbit calculations, and then proposed a fullerene-type $Si_{24}$-based configuration for Si nanowires. A part composed of pentagons in the cage of Fig. 2(b) is similar to the cages modeled by Marsen and Sattler [19]. Our results that Si nanowires applied to the PBC have included cages, are compatible with the fullerene-structured nanowires of Si modeled by $Si_N$-cage polymer structures.

Below results shows the tubular structures obtained from the SA simulations. Figures 6(a) and 6(b) show the final structure of the SA simulation with 1 % by 5 ps from 800 K for B1 case. Figure 6(a) shows two tubes connected with a common boundary composed of three Si atoms. However, since this structure is not well-defined structure, when three Si atoms, indicated by three arrows in Figs. 6(a) and 6(b), are manipulated, a well-defined SiNT structure is obtained as shown in Figs. 6(c) and 6(d). In Fig. 6(d), the dark and the bright spheres indicate $sp^3$ and $sp^2$ bonds, respectively. Si atoms in the common connected regions have $sp^3$ bonds. Figure 6(e) shows the cohesive energy per atom and the structural transitions as a function of temperature for the structure of Fig. 6(c). The structure of Fig. 6(c) maintained the original structure below 580 K, the structure was changed into a tubular structure from 600 to 730 K, and the tubular structure was spread out and transformed into a graphitelike sheet above 780 K.

Figure 7 shows a tube structure composed of a core atomic strand and an outer wall. An unit structure of Fig. 7 is a part of the final structure obtained from the SA simulation with quenching rate 1 % by 5 ps from 800 K for C1 case. Figures 7(a) and 7(b) shows the cross-sectional and side views and Figs. 7(a) and 7(b) show the core atomic strand and the outer wall, respectively. Where the form of a chemical bond between an atom in the core atomic strand and an atom on the outer wall, the atoms in the core have $sp^3$ bonds and the atoms on the outer wall have $sp^2$ bonds. This structure is unstable. While this structure was stabilized by the SD method at T = 0 K, this structure did not even maintain in MD simulations below 300 K.  While carbon is flexible with the type of hybridization ($sp^3$, $sp^2$, and $sp^1$), silicon is restricted to $sp^3$ and can also have $sp^2$ in the Si-based fullerene family including both $sp^3$ and $sp^2$ [18,19]. This structure including $sp^1$ seems to be achieved from a poor description of the Tersoff potential for ultrathin Si nanostructures.

Figures 6 and 7 shows that the chemical bonds in core regions consist of $sp^3$ bonds, whereas the chemical bonds on surface regions consist of $sp^2$ or $sp^1$ bonds. This result is partially in accordance with the results investigated by Menon and Richter [18] that showed some stable SiNTs which their geometries consisted of a core of fourfold coordinated atom surrounded by a threefold coordinated outer surface incorporating one of the most stable reconstruction of bulk Si.

Figure 8 shows the variation of the cohesive energy per atom of SiNTs corresponding to CNTs as a function of temperature and the structural transitions of (8, 8) SiNT for temperature. The structures of SiNTs have been obtained from the SD method in Sec. 2. The disintegration of SiNTs is clearly identified by the abrupt jump in the internal energy curve such as the curve of $Si_{60}$ in Fig. 1. Though the diameter of SiNT increases, the disintegration temperatures are almost constant, 1200 K, and are hardly related to the diameter of SiNT, because the disintegration processes are related to the interactions between neighbor atoms, the short-range interactions based on the Tersoff potential. (8, 8) SiNT has maintained the tubular structure below the disintegration temperature. Above the disintegration temperature, Si atoms tend to agglomerate in themselves toward an amorphous form. For Si, since $sp^3$ bond is more stable than $sp^2$ bond, during the disintegration, while the number of $sp^3$ bonds increases, the number of $sp^2$ bond decreases [23].

That the cohesive energies per atom for $sp^2$ are higher than that for $sp^3$ implies the difficulty in producing a single walled SiNTs or graphitelike sheets. Our results of the SA simulations also show that the single walled SiNTs composed of only $sp^2$ bonds are very difficult but SiNTs composed of both $sp^2$ and $sp^3$ are relatively more stable. This study based on the Tersoff potential shows a few ultrathin Si structures related to fullerenes or nanotubes, as shown in Figs. 2-4 and 6. A study of Li *et al* [21] based on the FP-LMTO-MD showd that an armchair (3, 3) CNT is not stable and easily deformed. However, they also showed when the CNT was capped by the insertion of several atoms on both ends, the encaged structure was relatively stable. Most of our results also include cage-like structures as shown in Figs. 2 and 3.

Although the results of single walled SiNTs based on the Tersoff potential are quantitatively in good agreement with the previous results [18,21-23] based on the first principle calculations, the Tersoff potential for Si has been widely used to study Si bulk of diamomd structures, and the Tersoff potential for carbon has been applied to the prediction of new forms of carbon material and to the investigation of fullerenes and CNTs, we have a question about the scientific exactitude of the Tersoff potential for silicon material. The previous works [18,21-23] including this work have considered the bare Si nanostructures and have not shown in any experimental result. However, the experimental and a few theoretical works have shown the metal-encapsulating Si cage clusters and nanotubes [26,27,40,41]. Therefore, we think that the scientific exactitude of the Tersoff potential for ultrathin Si nanostructures requires further investigations. However, in this work, we do not investigate thus topic but discuss only a problem about the bond-type transitions. From MD simulations of the heating and the quenching of B1 case, we investigate the structural transition, the diamond – graphitelike sheet transition, of ultrathin Si nanostructure.

Figure 9 shows the cohesive energy curves of both the heating and the quenching cases and includes four structures during the heating case and the final structure of the quenching case. The heating and the quenching rates are 1 K by 5 ps from 300 to 800 K and 1 K by 5 ps from 800 to 100 K, respectively. In the heating case, the original structures maintained below 500 K, a tubular structure was formed from 520 to 670 K, the tubular structure was disintegrated at near 690 K, and a graphitelike sheet was formed above 750 K. Above 670 K, the cohesive energy curve and the structures obtained from the quenching case are similar to those obtained from the heating case. However, at near 600 K, a cage structure was formed and maintained until 100 K during the quenching. The slopes in the cohesive energy curve for a tubular structure found during the heating are similar to those for a cage structure found during the quenching. The final structure of the quenching case includes five $sp^3$ bonds. During the heating, the diamond – tube and the tube – graphitelike sheet transitions were achieved. However, during the quenching, the graphitelike sheet – tube transition was achieved but the tube – diamond transition was not achieved. Since this difference can be induced by the large quenching rate 1 K/ 5 ps, we simulated a case of the less quenching rate 0.1 K / 5 ps. The SA simulation with the quenching rate 0.1 K/ 5 ps showed a encaged structure below 670 K but finally also did not show the fullerenelike cage - diamond transitions. Briefly, in the cases of MD simulations based on the Tersoff potential for ultrathin Si nanostructures, on heating, the $sp^3$ to $sp^2$ transitions are achieved, whereas on quenching, the $sp^2$ to $sp^3$ transitions are partially achieved. From this comparison, one can think that the Tersoff potential for Si do not efficiently describe the $sp^2$ to $sp^3$ transition. Therefore, while the Tersoff potential for Si quantitatively gives a good description of the well-ordered structures, such as diamond, $Si_{60}$ fullerene, and SiNTs corresponding to CNTs, this potential gives only a poor description of the structural transition of ultrathin Si nanostructures.

In our simulations based on the Tersoff potential, since the graphitelike sheet – tube transition has been often found, we investigated the energy barrier of the graphitelike sheet – tube transition. Figure 10 shows the energy diagram of the graphitelike sheet – tube transition for (5, 5) SiNT. Since the standard angle between the neighbor atom and the origin atom on the same layer is 30º for (5, 5) SiNT, the angle, $\theta_C$, increases to 30º by 1º with the fixed bond length, 2.305 Å. We calculated the structures of the curved-sheet corresponding to $\theta_C$ using both the angles between each atoms and $\theta_C$. The structure associated with $\theta_C$ was relaxed by the SD method on the condition that atoms of both ends were fixed, and the cohesive energies per atom were obtained from the relaxed structures. The total energy barrier of the graphite – tube transition is 0.020693 eV/atom. The first energy barrier ($E_1$) is 0.00981 eV/atom and the peak at 8º, and the stress to curve the sheet into tube increases from 13º and reaches the peak of the second energy barrier ($E_2$ = 0.015694 eV/atom) at 29º. As soon as the stress reaches the peak of the second barrier, the curved-sheet associated with 29º forms (5, 5) SiNT rapidly. This result, the low activation energy of the sheet – tube transition, are in excellent agreement with the results of the SA simulations that have shown encaged or tubular structures in most of the final structures obtained from the SA simulations in the work. The activation energy and the formation energy of the (5, 5) SiNT – graphite sheet transition are 0.12451 eV/atom and 0.103817 eV/atom, respectively. Considering that the kinetic energy per atom corresponding to 960 K is 0.125132 eV, the activation energy of the (5, 5) SiNT – graphite sheet transition is in concordance with MD simulation results that the disintegration temperatures of $Si_{60}$ fullerene and SiNTs are 1000 to 1200 K in Figs. 1 and 8

To conclude, we can analogize the whole out of this work as follows: When SiNTs are composed of both $sp^2$ and $sp^3$ bonds and the ratio of $sp^3$ to $sp^2$ is high, the probability of the existence of SiNTs increases. This interpretation is in good agreement with the results investigated by Menon and Richter [18] showing that the geometries of some stable SiNTs consist of a core of fourfold coordinated atom surrounded by a threefold coordinated outer surface and by Marsen and Sattler [19] proposing $Si_N$ cage polymer structures. Ultimately, ultrathin nanostructures consisting of Si atoms can be found in nanocages and nanotubes, because hollow region can be found in their cores in order to minimize the number of $sp^2$ bonds. Though the Tersoff potential for Si quantitatively gives a good description of the well-ordered structures, such as diamond, $Si_{60}$ fullerene, and SiNTs corresponding to CNTs, the Tersoff potential gives a poor description of the structural transition of ultrathin Si nanostructures.

### 4. Conclusion

This study showed the structures and the thermal behaviors of Si nanocages and nanotubes using classical molecular dynamics simulations based on the Tersoff potential. Si-Si bond length, cohesive energies per atom, diameters, and elastic energy to curve the sheet into tube of hypothetical Si nanotubes based on the Tersoff potential were in good agreement with those obtained from previous DFT results [22]. Most of the structures, which obtained from the SA simulations for several initial structures with diamond structure, have included encaged, tubular, or sheet-like structures and have been composed of both $sp^3$ and $sp^2$ bonds. A systematic study about the thermal behavior of a $Si_{60}$ fullerene and several hypothetical silicon nanotubes was presented. Though the diameter of silicon nanotubes increased, their disintegration temperatures were almost constant because the disintegration processes were mainly related to the interactions between neighbor atoms. The cohesive energies per atom for silicon nanotubes were higher than that for the Si bulk in the diamond structure, and this implies the difficulty in producing silicon nanotubes or graphitelike sheets. However, since the elastic energy per atom to curve the sheet into tube for silicon atoms is very low, if graphitelike sheets of silicon are formed, the extra cost to produce the tubes is also very low. When silicon nanotubes are composed of both $sp^2$ and $sp^3$ bonds and the ratio of $sp^3$ to $sp^2$ is high, the probability of the existence of silicon nanotubes increases. Therefore, silicon nanotubes similar to multi walled structures are anticipated. However, since this work is dependent on the scientific exactitude of the Tersoff potential, the fact that the Tersoff potential gives a poor description of the structural transition leaves further works to be desired.

**TABLE**

Table 1. Initial structure, diameter, number of atoms, and length of PBC of ultrathin Si nanowires.

|    | Initial structure | Diameter (Å) | Number of atoms | Length of PBC |
|----|-------------------|--------------|-----------------|---------------|
| A1 | {111}             | 7.8          | 52              | 17.24416      |
| A2 | {111}             | 16.2         | 220             | 17.24416      |
| A3 | {111}             | 22.1         | 364             | 17.24416      |
| B1 | {110}             | 9.0          | 45              | 11.51983      |
| B2 | {110}             | 16.7         | 132             | 11.51983      |
| B3 | {110}             | 23.5         | 255             | 11.51983      |
| C1 | {100}             | 8.6          | 63              | 16.29151      |
| C2 | {100}             | 15.9         | 171             | 16.29151      |
| C2 | {100}             | 22.4         | 327             | 16.29151      |

**FIGURES**

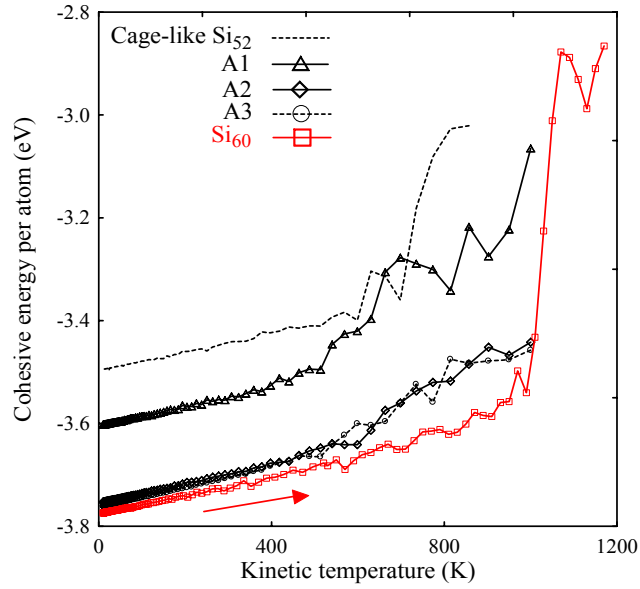

Figure 1. Cohesive energy per atom as a function of temperature for the SA simulations of A1, A2, and A3 with the quenching rate 5 % by 5 ps. Cage-like $Si_{52}$ was obtained from the SA simulation of A1 with the quenching rate 5 % by 5 ps without the PBC. The kinetic temperature of $Si_{60}$ fullerene increased by the heating rate 5 % from 10 to 400 K and by 20 K interval from 400 to 1200 K.

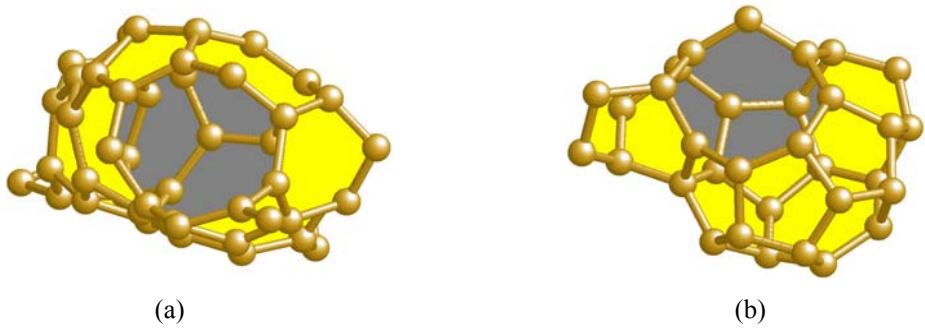

(a)                            (b)

Figure 2. (a) Cage-like $Si_{52}$ obtained from the SA simulation of A1 with the quenching rate 5 % by 5 ps without the PBC. (b) Cage-like Si32 that was extracted from the final structure, which was obtained from the SA simulation with the quenching rate 5 % by 5 ps from 1000 K for Al with the PBC.

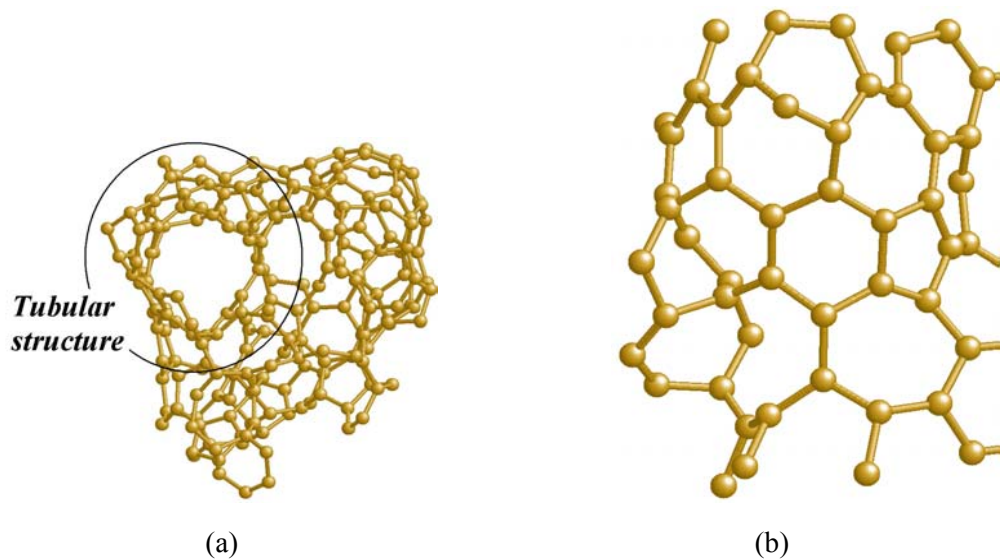

(a)                        (b)

Figure 3. (a) Final structure of the SA simulation with the quenching rate 5 % by 5 ps from 1000 K for A2 that includes a tubular and a few encaged structures. The circle indicates the tubular structure. (b) Sheet-like structure of the outside wall of the tubular structure indicated by the circle.

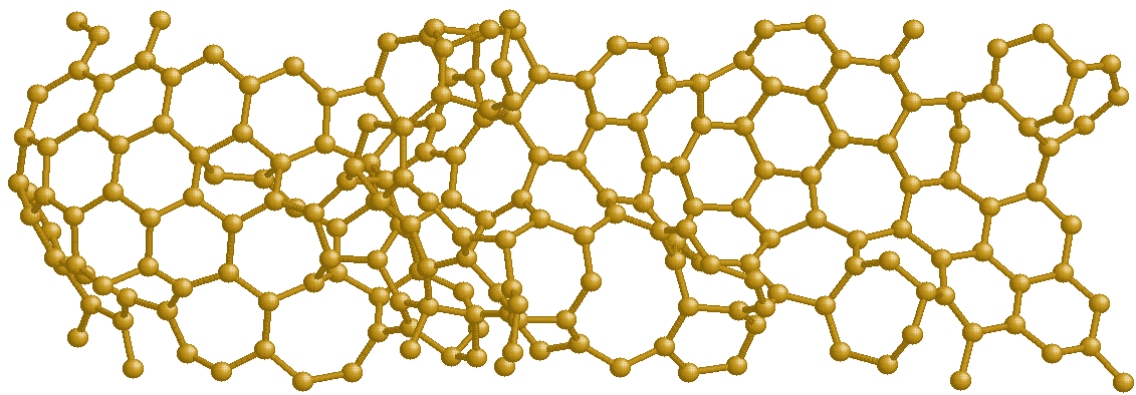

Figure 4. Final structure of the SA simulation with the quenching rate 1 % by 5 ps from 1200 K for A2.

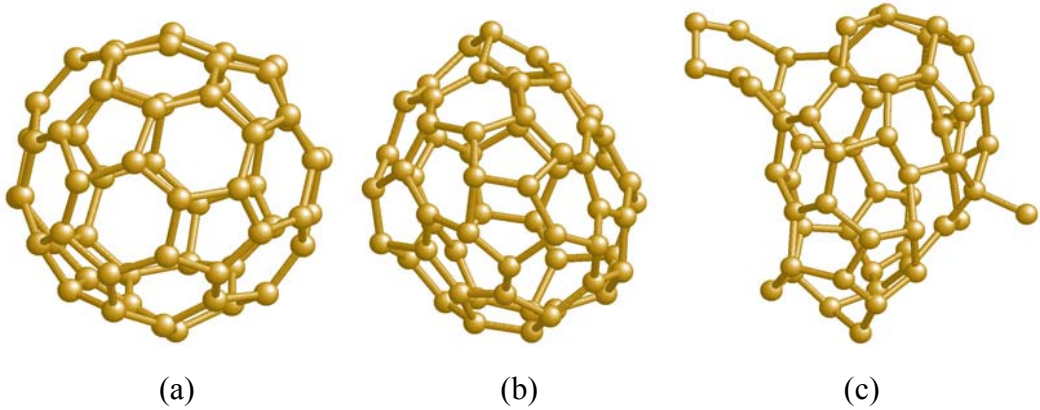

Figure 5. Structures of $Si_{60}$ fullerene for temperature. (a) 450 K, (b) 990 K, and (c) 1030 K.

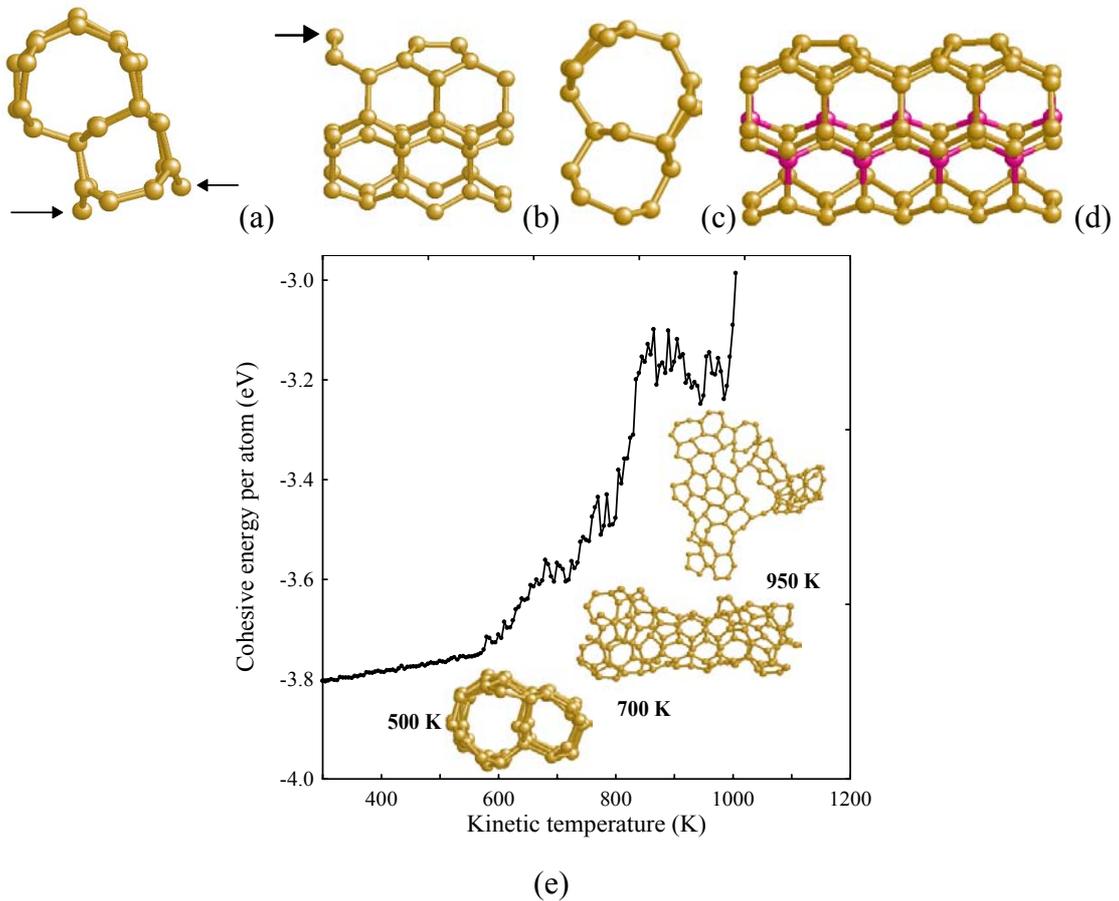

Figure 6. (a) Top and (b) side views of the final structure of the SA simulations with the quenching rate 1 % by 5 ps from 800 K for B1. Three arrows indicate the manipulated atoms to be a well-ordered structure. (c) Top and (d) side views of a well-ordered structure. Dark and bright spheres in (d) indicate $sp^3$ and $sp^2$ bonds, respectively. (e) Cohesive energy per atom as a function of temperature and structural transition for temperature.

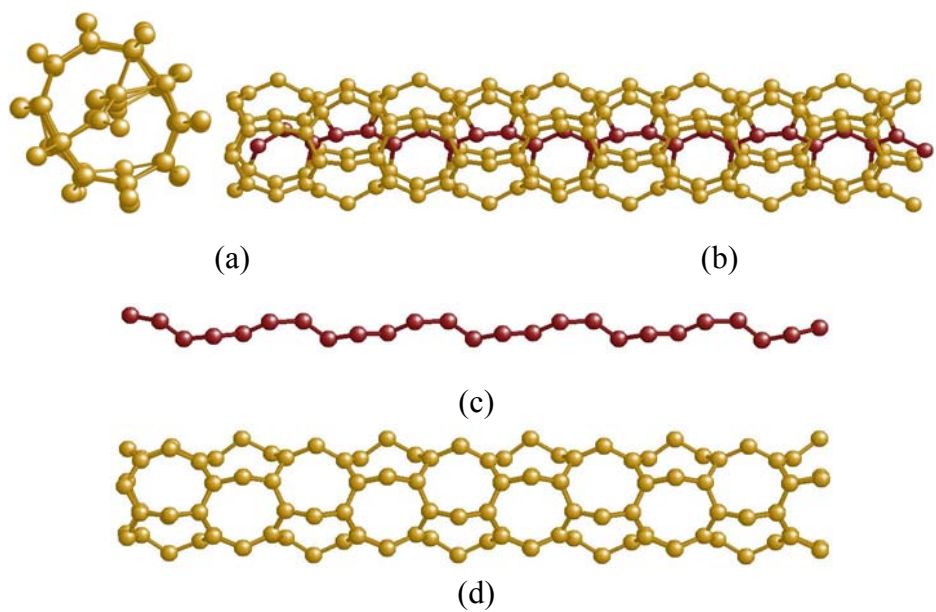

Figure 7. Tube structure composed of a core atomic strand and an outer wall. (a) Top and (b) side views. (c) Core and (d) outer wall structures.

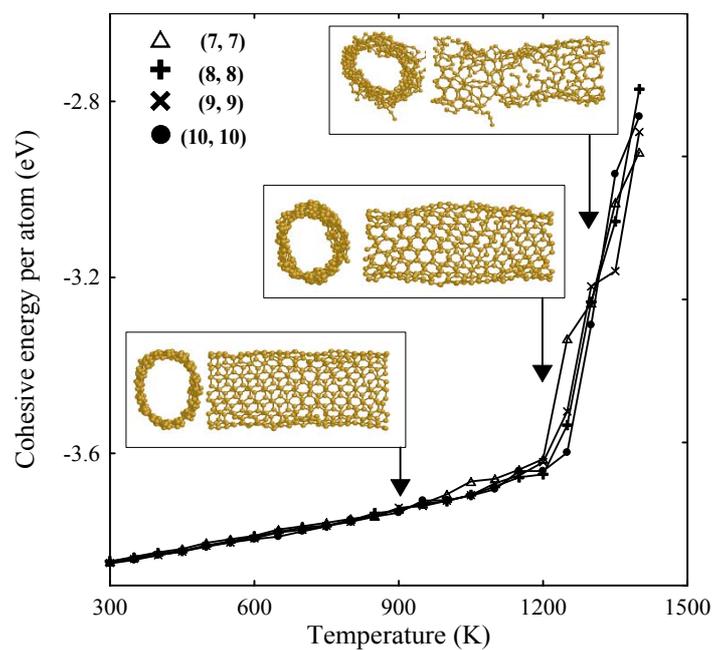

Figure 8. Cohesive energy per atom as a function of temperature for some (n, n) SiNTs corresponding to CNTs and the structural transition of (8, 8) SiNT for temperature.

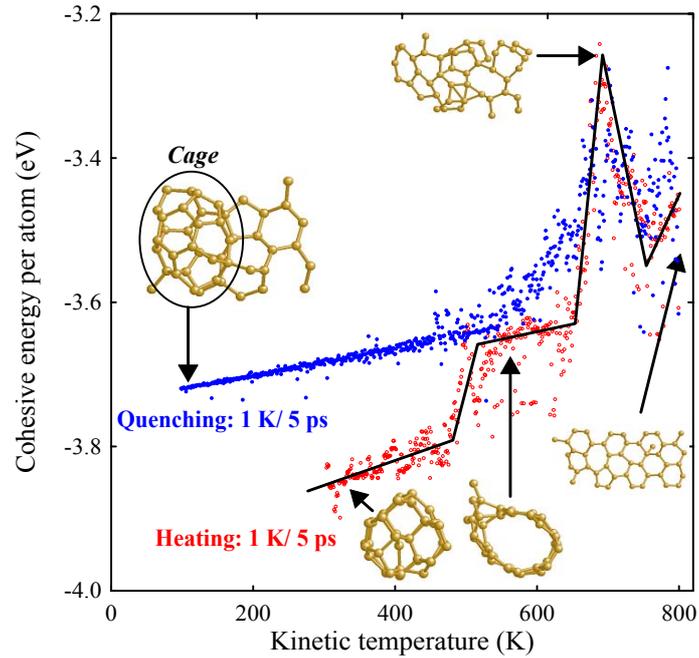

Figure 9. Cohesive energy curves as a function of temperature for both the heating and the quenching cases for B1. The heating and the quenching rates are 1 K by 5 ps from 300 to 800 K and 1 K by 5 ps from 800 to 100 K, respectively. Four structures during the heating cases and final structure of the quenching caes.

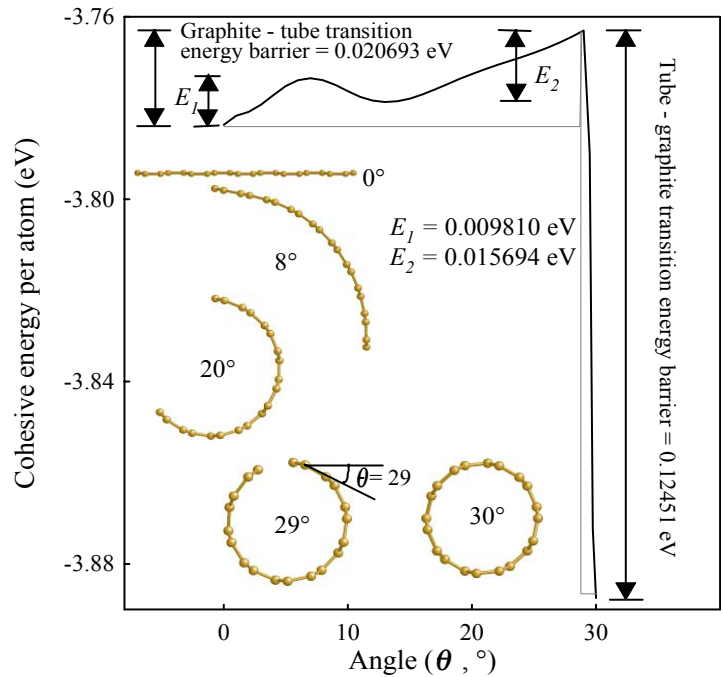

Figure 10. Energy diagram of the graphitelike sheet – nanotube transition for (5, 5) SiNT.